\documentclass[twocolumn,aps,prd,10pt,superscriptaddress,longbibliography]{revtex4-1}
\usepackage{amsmath}
\usepackage{bm}
\usepackage{glossaries}
\usepackage{graphicx}
\usepackage{multirow}
\usepackage{gensymb}
\usepackage{soul}
\usepackage[normalem]{ulem}
\usepackage[usenames,dvipsnames,svgnames]{xcolor}
\usepackage{hyperref}
\hypersetup{
pdfnewwindow=true,      % links in new window
colorlinks=true,        % false: boxed links; true: colored links
linkcolor=Blue,         % color of internal links
citecolor=Blue,         % color of links to bibliography
filecolor=Blue,         % color of file links
urlcolor=Blue           % color of external links
}

\newacronym{md}{MD}{molecular dynamics}
\newacronym{mlp}{MLP}{machine-learned potential}
\newacronym{gap}{GAP}{Gaussian-approximation potential}
\newacronym{soap}{SOAP}{smooth overlap of atomic positions}
\newacronym{nep}{NEP}{neuroevolution potential}
\newacronym{nn}{NN}{neural network}
\newacronym{snes}{SNES}{separable natural evolution strategy}
\newacronym{dft}{DFT}{density-functional theory}
\newacronym{rmse}{RMSE}{root-mean-square error}
\newacronym{nemd}{NEMD}{non-equilibrium molecular dynamics}
\newacronym{hnemd}{HNEMD}{homogeneous non-equilibrium molecular dynamics}
\newacronym{mfp}{MFP}{mean free path}
\newacronym{vdos}{VDOS}{vibrational density of states}
\newacronym{bte}{BTE}{Boltzmann transport equation}
\newacronym{ald}{ALD}{anharmonic lattice dynamics}
\newacronym{gpu}{GPU}{graphics processing units}
\newacronym{cpu}{CPU}{central processing unit}
\newacronym{cn}{CN}{coordination number}
\newacronym{pcf}{PCF}{pair-correlation function}
\newacronym{adf}{ADF}{angular-distribution function}
\newacronym{asi}{a-Si}{amorphous silicon}
\newacronym{sm}{SM}{Supplementary Material} % change this according to the journal
\newacronym{cvd}{CVD}{chemical vapor deposition}
\newacronym{pvd}{PVD}{physical vapor deposition}

\begin{document}

%\title{Temperature-Dependent Thermal Conductivity for Amorphous Silicon: Machine Learning Potential Driven Molecular dynamics calculation}
%%\textcolor{red}{\st{Quantum-corrected}} 

\title
{Quantum-corrected thickness-dependent thermal conductivity in amorphous silicon predicted by machine-learning molecular dynamics simulations}

\author{Yanzhou Wang}
\affiliation{Beijing Advanced Innovation Center for Materials Genome Engineering, Department of Physics, University of Science and Technology Beijing, Beijing 100083, China}
\affiliation{Department of Applied Physics, QTF Center of Excellence, Aalto University, FIN-00076 Aalto, Espoo, Finland}

\author{Zheyong Fan}
\email{brucenju@gmail.com}
\affiliation{Department of Applied Physics, QTF Center of Excellence, Aalto University, FIN-00076 Aalto, Espoo, Finland}
\affiliation{College of Physical Science and Technology, Bohai University, Jinzhou, 121013, China}

\author{Ping Qian}
\email{qianping@ustb.edu.cn}
\affiliation{Beijing Advanced Innovation Center for Materials Genome Engineering, Department of Physics, University of Science and Technology Beijing, Beijing 100083, China}

\author{Miguel A. Caro}
%\email{mcaroba@gmail.com}
\affiliation{Department of Electrical Engineering and Automation, Aalto University, FIN-02150 Espoo, Finland}
\affiliation{Department of Chemistry and Materials Science, Aalto University, FIN-02150 Espoo, Finland}

\author{Tapio Ala-Nissila}
\email{tapio.ala-nissila@aalto.fi}
\affiliation{Department of Applied Physics, QTF Center of Excellence, Aalto University, FIN-00076 Aalto, Espoo, Finland}
\affiliation{Interdisciplinary Centre for Mathematical Modelling and Department of Mathematical Sciences, Loughborough University, Loughborough, Leicestershire LE11 3TU, United Kingdom}

\date{\today}

%keywords. Machine learned neuroevolution potential, amorphous silicon, molecular dynamics, thermal conductivity

\begin{abstract}
Amorphous silicon (a-Si) is an important thermal-management material and also serves as an ideal playground for studying heat transport in strongly disordered materials. Theoretical prediction of the thermal conductivity of a-Si in a wide range of temperatures and sample sizes is still a challenge.  Herein we present a systematic investigation of the thermal transport properties of a-Si by employing large-scale molecular dynamics (MD) simulations with an accurate and efficient machine-learned neuroevolution potential (NEP) trained against abundant reference data calculated at the quantum-mechanical density-functional-theory level. The high efficiency of NEP allows us to study the effects of finite size and quenching rate in the formation of a-Si in great detail. We find that a simulation cell up to $64,000$ atoms (a cubic cell with a linear size of 11 nm) and a quenching rate down to $10^{11}$ K s$^{-1}$ are required for almost convergent thermal conductivity.  Structural properties, including short- and medium-range order as characterized by the pair correlation function, angular distribution function, coordination number, ring statistics and structure factor are studied to demonstrate the accuracy of NEP and to further evaluate the role of quenching rate. Using both the heterogeneous and homogeneous nonequilibrium MD methods and the related spectral decomposition techniques, we calculate the temperature- and thickness-dependent thermal conductivity values of a-Si and show that they agree well with available experimental results from 10 K to room temperature. Our results also highlight the importance of quantum effects in the calculated thermal conductivity and support the quantum-correction method based on the spectral thermal conductivity.
\end{abstract}

\maketitle

\section{Introduction}

Silicon remains as one of the most fundamental semicondutor materials in the microelectronics industry. \Gls{asi} is a disordered semiconductor material with important technological
applications, in particular as photoabsorber in solar cells~\cite{2017_Stuc_a-Si-solarCell,2020_Rama_a-Si-solarCell}. 
Understanding the thermal properties of \gls{asi} at the atomic level is important in predicting the behavior of this material and how it may affect device performance.
Experimental measurements \cite{1994_Cahill_kappa-expt, 2006_prl_zink, regner2013nc, 2021_prm_Kim_kappa-expt} have played an important role in characterizing the thermal transport properties of \gls{asi}, but theoretical understanding and reproduction of the experimental results are also important. For pristine crystalline silicon, phonons are the dominant heat carriers and phonon-mediated heat transport has been well understood in terms of anharmonic phonon-phonon scattering within the phonon-gas picture. However, due to the complexity of the structure and the absence of long-range order \cite{1991_nature_Elliott} in amorphous systems, there are no well-defined phonon bands and the phonon-gas picture is not valid \cite{lv2016sr}. Most of heat carriers in \gls{asi} are vibrations with short \glspl{mfp} due to the disorder-induced scattering and thus the thermal conductivity in \gls{asi} is about two orders of magnitude smaller than that in crystalline silicon around room temperature \cite{regner2013nc}.

Assuming the dominance of disorder-mediated scattering, a harmonic Hamiltonian model has been proposed by Allen and Feldman \cite{allen1993prb} and numerical methods based on the Kubo-Greenwood formula have been devised \cite{Feldman1993prb}, which have led to a classification of lattice vibrations in amorphous materials into propagons, diffusons, and locons corresponding to low, medium, and high-frequency vibrations, respectively \cite{Allen1999pmb}. 
Recently, unified approaches that can account for both anharmonicity and disorder have been developed \cite{isaeva2019nc,simoncelli2019np}, providing a more comprehensive understanding of heat transport from the crystalline to the strongly disordered limit. These methods have found many applications in amorphous or amorphous-like materials, yet they have limitations, such as nonlinear scaling of the computational cost with respect to the simulation cell size and the high cost of including high-order anharmonicity.

\Gls{md} simulations, on the other hand, have a linear-scaling computational cost with respect to the simulation cell size and contain lattice anharmonicity and phonon scatterings to all orders. MD is the most comprehensive classical atomistic simulation method to study thermal transport and has been the standard approach used for benchmarking other theoretical models or computational methods \cite{isaeva2019nc, 2021_zhou_jap, zhang2022npj}.
However, reliable application of \gls{md} simulations to amorphous materials in general, and to \gls{asi} in particular, is hindered by two main aspects: the scarcity of accurate and efficient interatomic potentials and the classical nature of the \gls{md} method.

In this paper, we present solutions to both of the aforementioned obstacles. On the one hand, we develop an accurate yet highly efficient interatomic potential based on machine-learning techniques for general silicon systems, applicable to \gls{asi} in particular. There have been some \glspl{mlp} developed for studying thermal transport in \gls{asi} \cite{2019_mtp_qian, 2020_mtp_Li}, but they are not efficient enough for performing a comprehensive investigation with careful convergence tests. The silicon \gls{mlp} we develop in this work is based on the \gls{nep} framework \cite{2021_prb_fan} which can achieve an unprecedented computational speed of about $10^7$ atom-step per second using a single Nvidia \gls{gpu} such as Tesla A100. The high accuracy and efficiency of the NEP model allow us to reach a large simulation cell size and long evolution times, generating realistic \gls{asi} structures that closely resemble the experimental samples, which is the prerequisite for obtaining reliable predictions for the thermal transport properties. On the other hand, we apply a proper quantum-statistical correction to the spectral thermal conductivity calculated within the \gls{hnemd} formalism \cite{2019_prb_fan_hnemd} and find that this can lead to quantitative agreement with the experimental results \cite{2006_prl_zink} in a wide range of temperatures, from 10 K up to the room temperature. With the combination of the efficiency and accuracy of the \gls{nep} model and a proper quantum-statistical correction, we achieved insightful results that are difficult to obtain previously.

\section{Training a machine-learned potential for a-Si}

\begin{figure}
    \centering
    \includegraphics[width=\columnwidth]{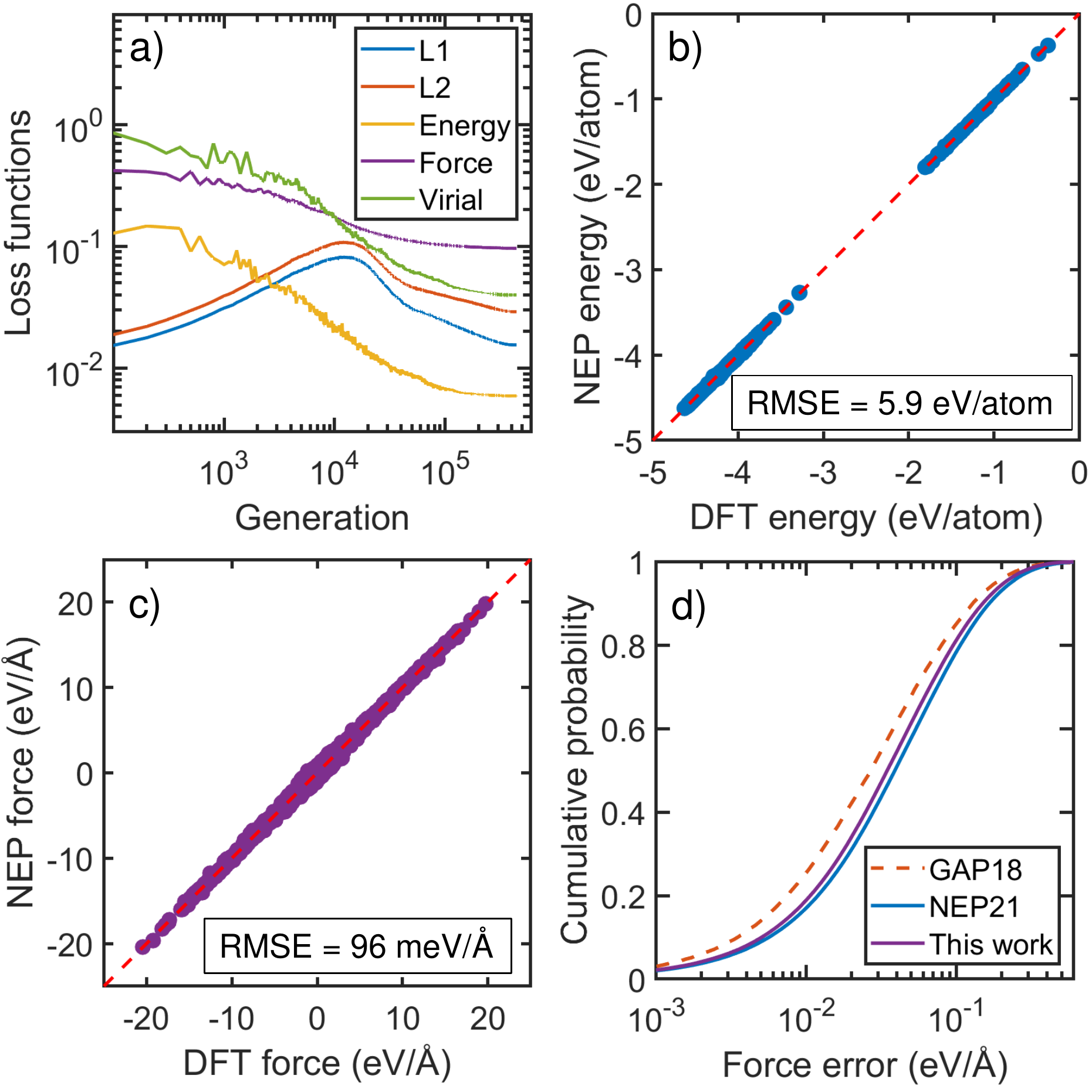}
    \caption{(a) Evolution of the various terms in the loss function, including those for the $\mathcal{L}_1$ and $\mathcal{L}_2$ regularization, the energy root-mean-square error (RMSE) (eV/atom), force RMSE (eV/\AA) and virial RMSE (eV/atom) as a function of the training generation. (b) Energy and (c) force  calculated from the \gls{nep} model as compared to the PW91-\gls{dft} reference data. The overall converged RMSE of energy and force are presented. (d) Cumulative probability of force error from the \gls{nep} model trained in this work as compared to those from the previous NEP21~\cite{2021_prb_fan} and  GAP18~\cite{2018_prx_Bartok} models.}
    \label{fig:nep}
\end{figure}

The \gls{nep} model is a \gls{nn} based \gls{mlp} trained using the \gls{snes}~\cite{Schaul2011}. The NN maps the local atom-environment descriptor of a central atom to its site energy and the total energy of an extended system is the sum of the individual site energies of the atoms. The descriptor used in \gls{nep}~\cite{2021_prb_fan} consists of selected radial and angular components similar in spirit to the Behler-Parrinello symmetry functions~\cite{2011_Behler_jcp} and the optimized~\cite{2019_Miguel_soap} \gls{soap}~\cite{bartok_2013}.

To train a \gls{nep} model applicable to \gls{asi}, we reuse the well-designed training database that has been used to train an accurate \gls{gap} (called GAP18 here)~\cite{2018_prx_Bartok}. This training dataset was computed at the quantum-mechanical density-functional (\gls{dft}) level using the PW91 functional~\cite{1992_Perdew_pw91_prb} and covers a wide range of silicon structures, including the liquid and \gls{asi} ones in particular. The GAP18 potential has been demonstrated to be well transferable and is able to simultaneously describe various properties of crystalline and non-crystalline silicon~\cite{2018_jpcl_Deringer}. However, thermal transport usually involves large length and long time scales and GAP18 is not currently efficient enough for this purpose. The NEP model as implemented in the \textsc{gpumd} package~\cite{2017_fan_gpumd,fan2022GPUMD}, on the other hand, can reach a computational speed of about $5\times 10^6$ atom-step per second for \gls{asi} by using a single \gls{gpu} card such as Tesla V100, which is about three orders of magnitude faster than GAP18 using 72 Xeon-Gold 6240 \gls{cpu} cores~\cite{2021_prb_fan}. 

A \gls{nep} model has already been trained previously for bench-marking the NEP framework~\cite{2021_prb_fan} (we call it NEP21), but with a better understanding on the hyperparameters, we here re-train it by changing the relative weight of virial from $1$ to $0.1$, keeping all the other hyperparameters as used in Ref.~\onlinecite{2021_prb_fan} unchanged. Figure~\ref{fig:nep}(a) shows the convergence trend of the \gls{rmse} of energy, force, and virial during the training process. We note that both $\mathcal{L}_1$ and $\mathcal{L}_2$ regularization are used in our training, which can help to increase the robustness of the potential. Upon convergence, the predicted energy and force from NEP correlate with the reference data very well, as shown in Figs.~\ref{fig:nep}(b) and~\ref{fig:nep}(c). The converged energy and force \glspl{rmse}  for the training data set are $5.9$ meV/atom and $96$ meV/\AA{} respectively. The corresponding \glspl{rmse} for the hold-out testing data set as used in Ref.~\onlinecite{2018_prx_Bartok} are $7.8$ meV/atom and $93$ meV/\AA{} respectively. As can be seen in Fig.~\ref{fig:nep}(d), the NEP model trained in this work is slightly more accurate than NEP21 but is still less accurate than GAP18. This training accuracy is similar to that obtained by using the atomic cluster expansion approach~\cite{2022_Dusson_ace}. Despite the relatively lower training accuracy, the NEP trained here exhibits performance on a par with GAP18 in predicting the various structural properties of \gls{asi}, as will be demonstrated below.

\begin{figure}
    \centering
    \includegraphics[width=\columnwidth]{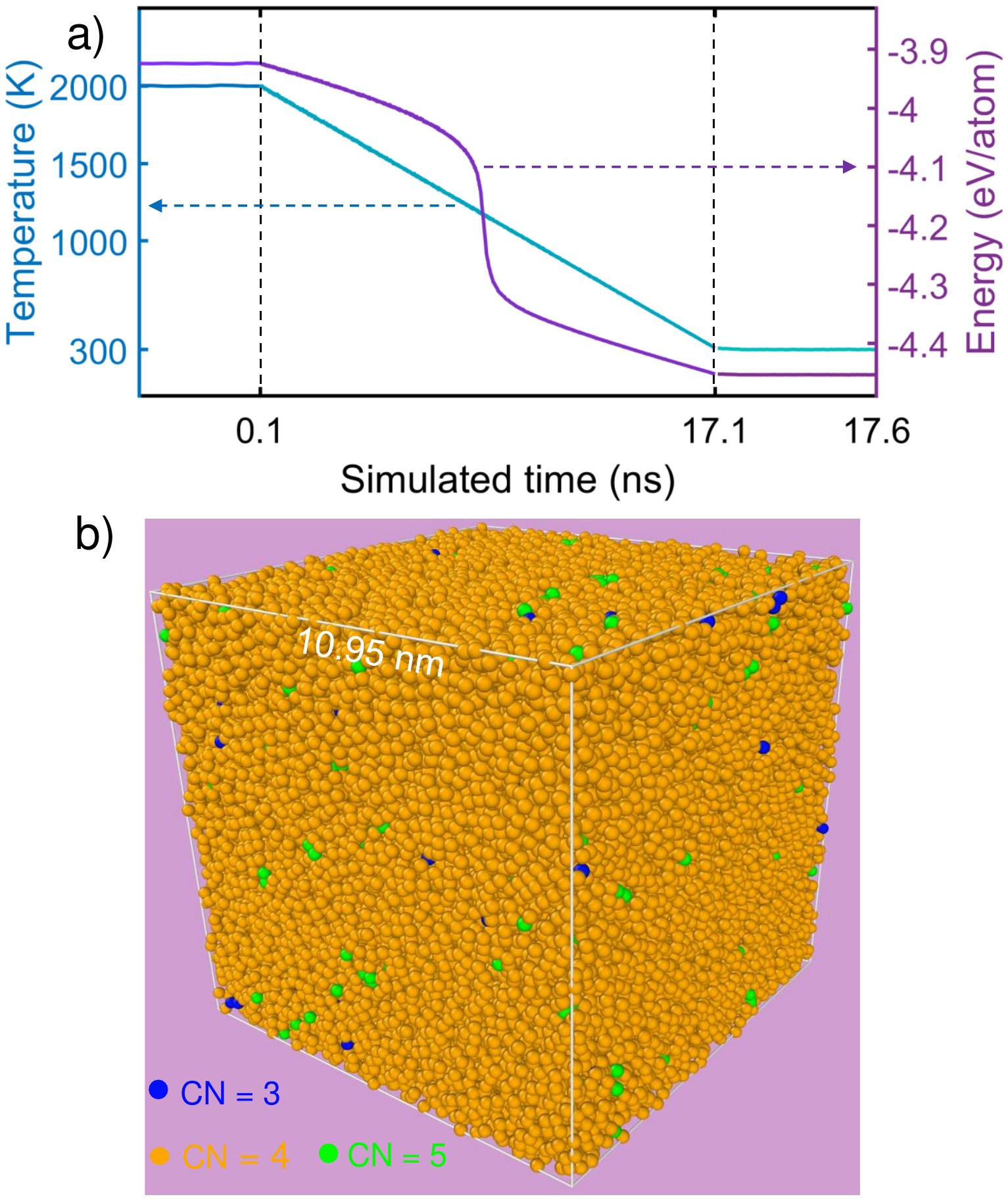}
    \caption{(a) Temperature and potential energy as a function of simulation time during the melt-quench-anneal process. (b) Snapshot of an \gls{asi} sample after the melt-quench-anneal process, where atoms with different coordination numbers (CN) are rendered in different colors. The \textsc{ovito} package~\cite{2010_ovito} is used for visualization.}
    \label{fig:temp_visualization}
\end{figure}

\section{Sample generation and structural characterization}

\subsection{Generating a-Si samples}

The key ingredient for obtaining reliable results for the physical properties of a-Si is sample preparation. To this end, we use classical \gls{md} simulations with a melt-quench-anneal process to prepare the \gls{asi} samples. All the \gls{md} simulations are performed using the \textsc{gpumd} package~\cite{2017_fan_gpumd} (version 2.9.1). We take diamond silicon as the initial structure and quickly heat it up to $T_0=2000$ K to reach the liquid state and equilibrate it for $0.1$~ns. Then, we cool down the system with the target temperature in the thermostat linearly dropping from $T_0$ to a temperature $T$ (from $10$ to $1000$ K) with a given  quenching rate $\alpha$ (from $5\times10^{12}$ down to $10^{11}$ K s$^{-1}$). Finally, we anneal the quenched sample at $T$ for $0.5$~ns to obtain a well equilibrated \gls{asi} structure. We use the isothermal-isobaric ensemble (zero target pressure) realized by the Berendsen thermostat and barostat~\cite{1984_berendsen} during the melt-quench-anneal process. We have checked that using the recently proposed Bernetti-Bussi barostat~\cite{2020_BBbarostat} combined with the Bussi-Donadio-Parrinello thermostat~\cite{2007_BDP_thermostat} does not lead to noticeably different results. In all the MD simulations, we use a time step of $0.5$~fs. 

Figure~\ref{fig:temp_visualization}(a) shows the evolution of the temperature and potential energy in the case of $T=300$ K and $\alpha=10^{11}$ K s$^{-1}$ obtained by using a system with $N=64,000$ atoms. In this case, the quenching process lasts $17$~ns. To appreciate the high computational demands for heat transport applications and the excellent computational efficiency of NEP we note that, to reach the same quenching rate, the system size must be chosen to be $N=512$ in GAP18~\cite{2018_jpcl_Deringer} with a time step of $1$~fs. 

\begin{figure}
    \centering
    \includegraphics[width=\columnwidth]{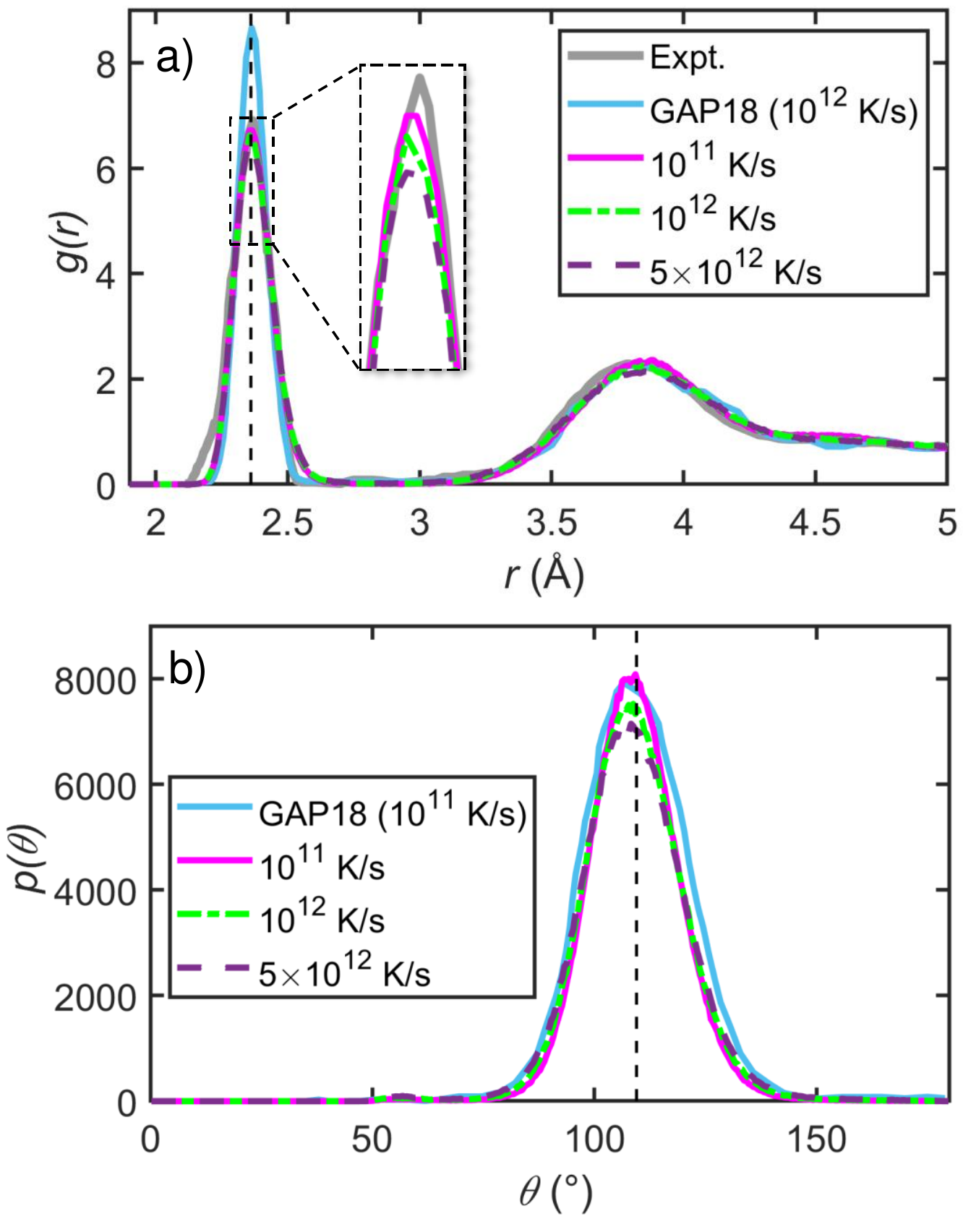}
    \caption{(a) Pair correlation function $g(r)$ and (b) angular distribution function $p(\theta)$ of $64,000$-atom \gls{asi} at $300$~K generated using different quench rates. The vertical dash lines in (a) and (b) mark the first peak of  $g(r)$ at $2.36$ \AA{} and that for $p(\theta)$ at $109.5 \degree$, respectively. Experimental measurements~\cite{1999_prl_Khalid,1999_prb_Khalid} and theoretical predictions for $g(r)$~\cite{2018_prx_Bartok} and  $p(\theta)$~\cite{2018_jpcl_Deringer} from GAP18 are presented for comparison. }
    \label{fig:rdf_adf}
\end{figure}

\subsection{Short-range order}

After generating the \gls{asi} samples, we first characterize the bond motifs of the short-range order in terms of the \gls{pcf}, the \gls{adf}, and the \gls{cn}. The results are shown in Fig.~\ref{fig:rdf_adf} and Table~\ref{tab:coordination}. For the first peak located at about $2.36$~\AA{} in the \gls{pcf}, the height calculated from the NEP model increases with decreasing quenching rate $\alpha$, getting close to the experimental value~\cite{1999_prl_Khalid} when $\alpha$ is reduced to $10^{11}$ K s$^{-1}$. The GAP18 model~\cite{2018_prx_Bartok} gives a sharper distribution around the first peak. All the theoretical and experimental results agree well beyond the first peak, particularly at the second peak at $3.86$~\AA{}. 

For the \gls{adf}, there are no experimental data, but the NEP and GAP18 models agree well for the same quenching rate $\alpha=10^{11}$~K s$^{-1}$. Both show a peak at an angle of $109.5\degree$, indicating the dominance of $sp^3$ bond motifs in \gls{asi}. Similar to the case of \gls{pcf}, the peak height in the \gls{adf} increases with decreasing quenching rate which indicates that a smaller quenching rate leads to a more locally ordered \gls{asi} structure.

\begin{table}[]
    \centering
    \setlength{\tabcolsep}{2Mm}
    \caption{Fractions of the different \glspl{cn} and the average \gls{cn} ($\overline{\rm{CN}}$) from our NEP model for different quenching rates. 
    }
    \begin{tabular}{lllll}
    \hline
    \hline
        $\alpha$ (K s$^{-1}$) & $\rm{CN}=3$ & $\rm{CN}=4$ & $\rm{CN}=5$ & $\overline{\rm{CN}}$ \\
    \hline
        $10^{11}$  & $0.45$\%  &  $98.09$\% & $1.46$\% & $4.010$ \\
        $10^{12}$  & $0.61$\%  &  $96.99$\% & $2.40$\% & $4.018$ \\
        $5\times10^{12}$ & $0.79$\%  &  $95.72$\% & $3.49$\% & $4.027$ \\
        $10^{11}$~\cite{2018_jpcl_Deringer} & $0.60$\%  & $98.36$\% & $1.04$\% & $4.004$ \\
        \hline
        \hline
    \end{tabular}
    \label{tab:coordination}
\end{table}

Based on the \gls{pcf}, we determine the \gls{cn} of each atom from the neighboring atoms within a cutoff distance of $2.9$ \AA{}. The calculated fractions of atoms with different \glspl{cn} and the average \gls{cn} are presented in Table~\ref{tab:coordination}. Most atoms have a \gls{cn} of 4 and the percentage of these atoms increases from $95.72\%$ to $98.09\%$ as the quenching rate decreases from $5\times10^{12}$~K s$^{-1}$ to $10^{11}$~K s$^{-1}$. This trend is in good agreement with that from GAP18~\cite{2018_jpcl_Deringer}. The results here again indicate that a lower quenching rate leads to a more locally ordered \gls{asi} sample. At the lowest quenching rate here,  the averaged \glspl{cn} from our \gls{nep} model and GAP18 are both close to 4. In contrast, the experimentally annealed \gls{asi} samples prepared by ion implantation have an averaged \glspl{cn} of $3.88$~\cite{1999_prb_Khalid}. This can be understood by noting that the experimental \gls{asi} samples are 1.8\% less dense than the crystalline precursor that has a \gls{cn} of 4 due to the appearance  of vacancy defects. 

\begin{figure}
    \centering
    \includegraphics[width=\columnwidth]{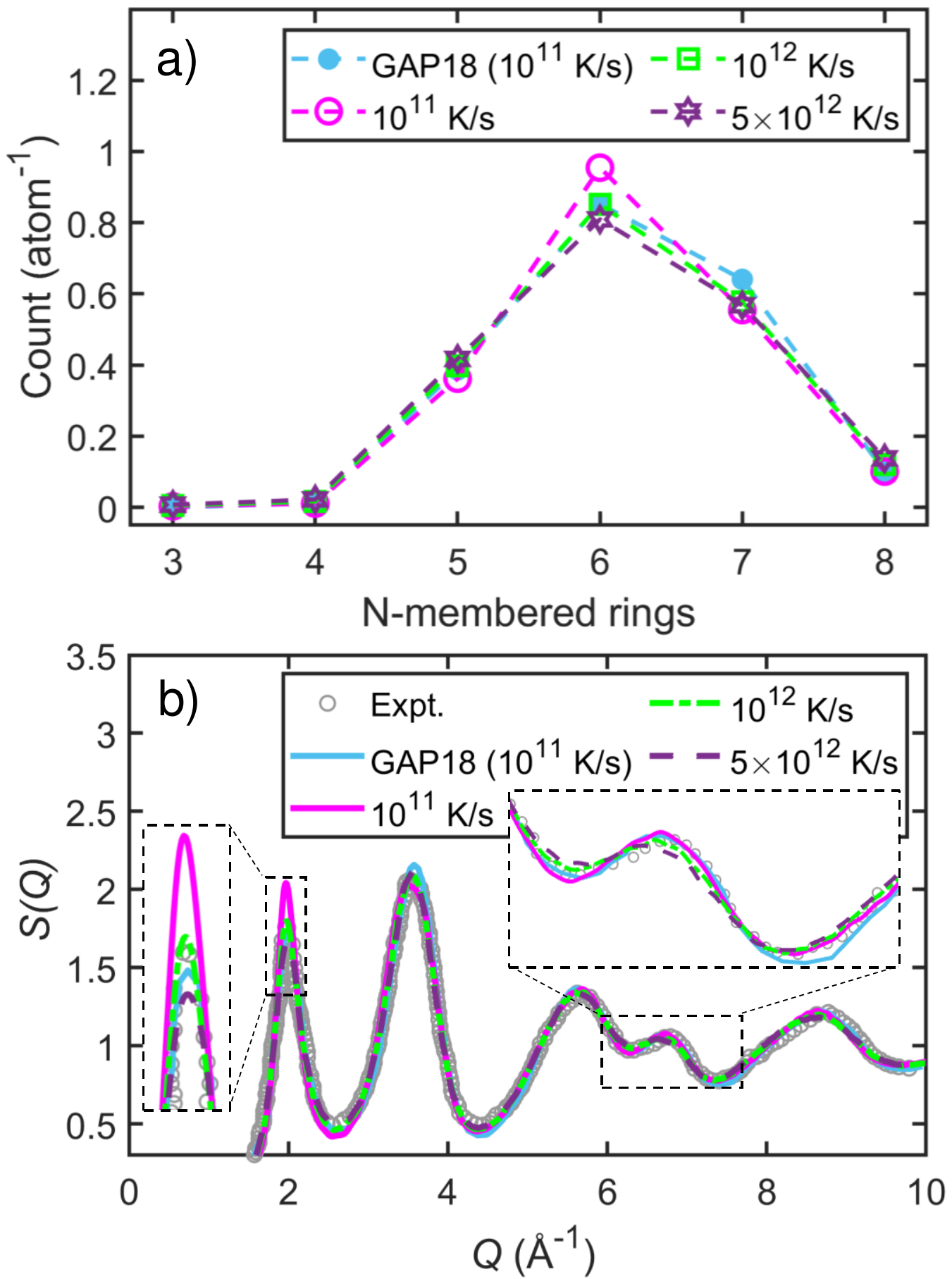}
    \caption{(a) Number of  $N$-membered rings per atom and (b) structure factor $S(Q)$ as a function of the wave vector $Q$ for \gls{asi} with 64,000 atoms at $300$~K. Experimental data~\cite{1999_prl_Khalid,1999_prb_Khalid} and prediction from GAP18~\cite{2018_jpcl_Deringer} are given for comparison.}
    \label{fig:ring_sq}
\end{figure}

\subsection{Medium-range order}
\label{section:Medium-range order}

Apart from the short-range order as characterized by the \gls{pcf}, \gls{adf}, and \gls{cn}, we also characterize the medium-range order. Ring motifs involve the sequential connections of coordination tetrahedra and can be used to characterize near or intermediate medium-range order that is the next length scale following the short-range domain \cite{1991_nature_Elliott}. We compute the ring distribution in Fig.~\ref{fig:ring_sq}(a) using shortest-path algorithm \cite{ring_fran_prb_1991}. For crystalline diamond silicon, all atoms are connected to one another in the cyclohexane-like 6-membered units with one ring per atom. For \gls{asi}, the most energetically favorable $6$-membered rings still dominate but 7- and 5-membered rings are also energetically viable and exist with a considerable amount as defected motifs in the three-dimensional network of \gls{asi}~\cite{2018_jpcl_Deringer}. With decreasing quench rate, the average number of 6-membered rings increases, which indicates an increased near or intermediate medium-range ordering. 

The structure factor $S(Q)$, one of the most common experimental structural probes, is typically regarded as a signature of medium-range ordering~\cite{1991_nature_Elliott,Hejna2013prb,Xie2013}. Computationally, $S(Q)$ is typically derived as the Fourier transform of the \gls{pcf}, i.e., $S(Q)=1+4\pi\rho\int_0^{\infty}r^2({\sin{Qr}}/{Qr})[g(r)-1]\, \text{d}r$. For the comparison with diffraction experiments, we also calculate the static $S(Q)$ in Fig.~\ref{fig:ring_sq}(b) using the \textsc{isaacs} package~\cite{2010_isaacs}. The first peak usually gives an indication of intermediate or far medium-range ordering~\cite{2018_jpcl_Deringer,Hejna2013prb,Xie2013}, and we see that it strengthens (in terms of increased $S(Q)$ value) at about $2$~\AA$^{-1}$ with decreasing quenching rate. Specifically for a-Si, there is a direct connection between the first peak in $S(Q)$ and the second peak in the \gls{pcf}, which corresponds to second-neighbors distances~\cite{dahal_2021}. Also, the shoulder peak at about $Q=7$~\AA$^{-1}$ features more clearly for lower quenching rate (see inset of Fig.~\ref{fig:ring_sq}(b)). Indeed, Laaziri \textit{et al.}~\cite{1999_prb_Khalid} observed experimentally that annealed \gls{asi} samples exhibit a more featured shoulder peak than as-deposited ones. The quantitative differences between the results from \gls{nep} and GAP18 as shown in Fig.~\ref{fig:ring_sq} are most likely due to the different simulation cell sizes.

\begin{figure}
\centering
\includegraphics[width=1\columnwidth]{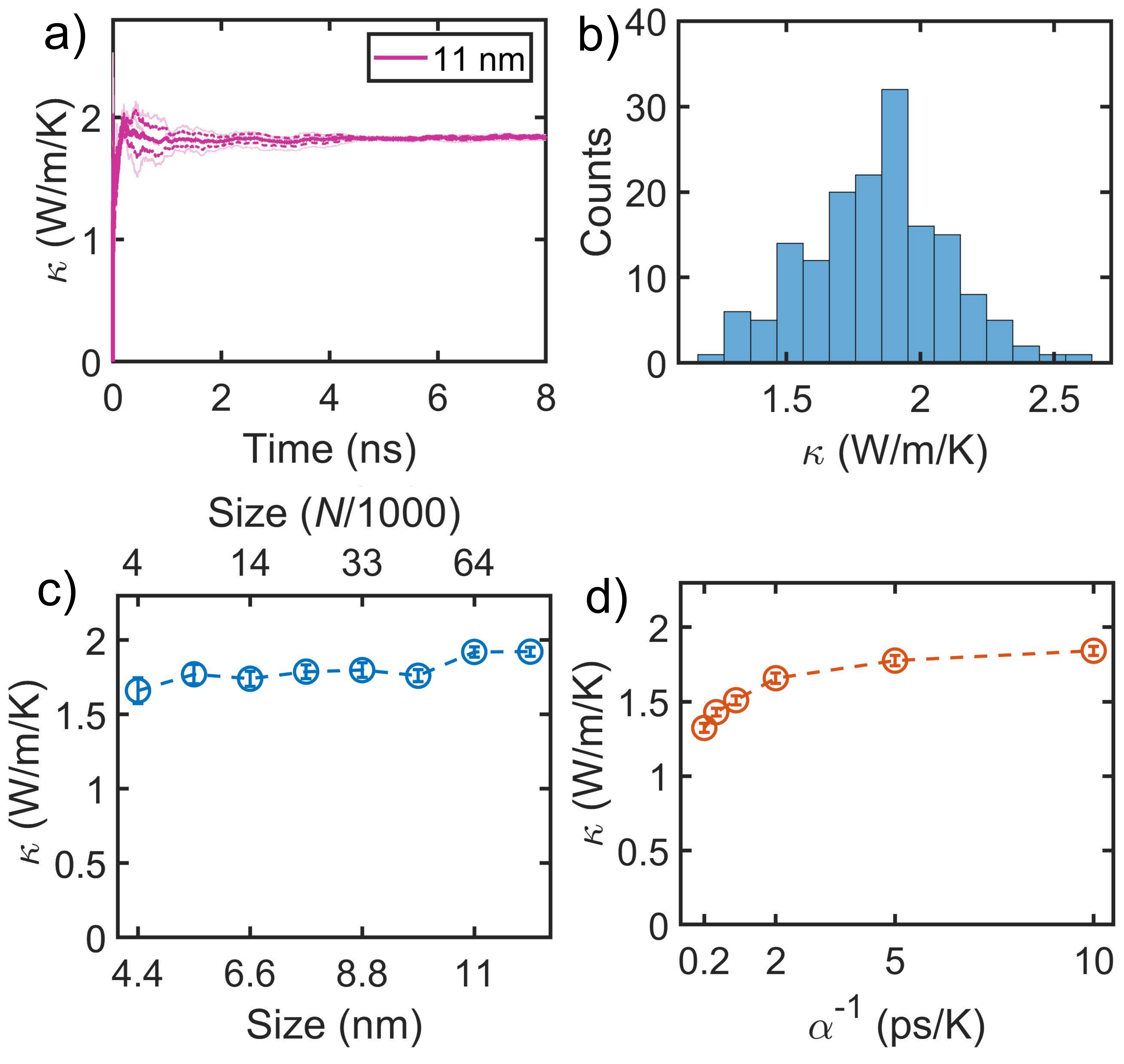}
\caption{(a) Cumulative average of the thermal conductivity $\kappa$ as a function of the \gls{hnemd} production time. The thin lines represent results from three independent \gls{asi} samples, and the thick and dashed lines represent the average and error bounds. In this case, the simulation cell size is $N=64,000$ and the quenching rate is $\alpha=10^{11}$~K s$^{-1}$. (b) Distribution of the block-averaged $\kappa$ values (each with 0.3 ns) from the HNEMD calculations. (c) $\kappa$ as a function of the \gls{asi} sample size (indicated as $N$ as well as the linear size of the cubic cell) with $\alpha=10^{11}$~K s$^{-1}$. (d) $\kappa$ as a function of inverse quenching rate $\alpha^{-1}$  with $N=64,000$. In all the cases, the temperature is $T=300$~K.} 
\label{fig:kappa-time-size-rate}
\end{figure}

\begin{figure*}[htb]
\centering
\includegraphics[width=1.5\columnwidth]{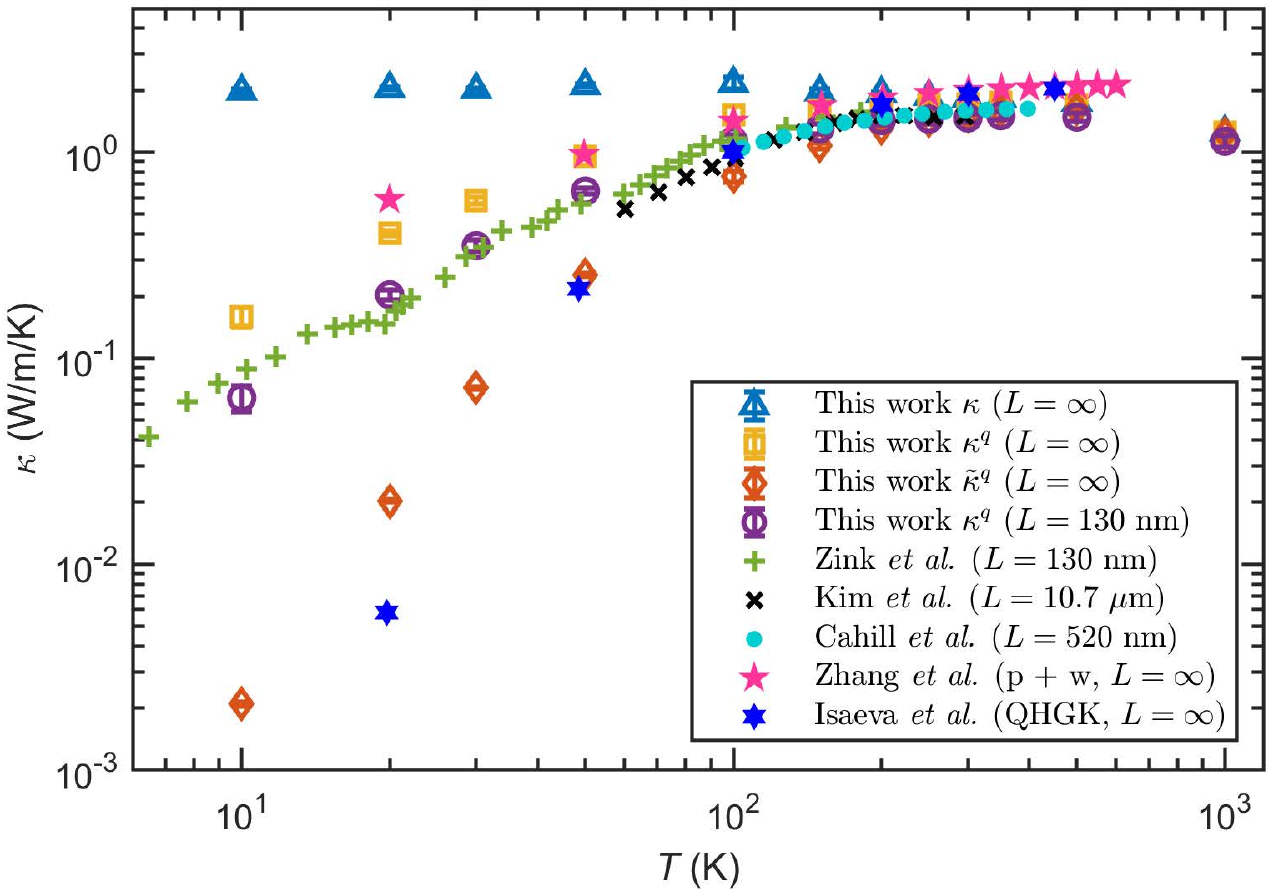}
\caption{Thermal conductivity $\kappa$ as a function of temperature $T$ for \gls{asi} from our calculations as compared to the experimental values by Zink \textit{et al.}~\cite{2006_prl_zink}, Kim \textit{et al.}~\cite{2021_prm_Kim_kappa-expt}, and Calhill \textit{et al.}~\cite{1994_Cahill_kappa-expt}. The quasi-harmonic Green-Kubo (QHGK) results by Isaeva \textit{et al.}~\cite{isaeva2019nc} and the theoretical results by Zhang \textit{et al.}~\cite{zhang2022npj} based on a particle-like and wave-like (``${\rm p + w}$'') decomposition are also shown for comparison. The sample thickness in the heat transport direction is indicated as $L$, where $L=\infty$ means the bulk limit.}
    \label{fig:kT}
\end{figure*}

\section{Heat transport in amorphous silicon}

\subsection{Effects of size and quenching rate on thermal conductivity}

After generating and thoroughly characterizing the structures of the \gls{asi} samples, we study their heat transport properties.
There are numerous methods for computing thermal conductivity at the atomistic level~\cite{2021_Guxiaokun_jap}, but \gls{md} in particular is efficient for strongly disordered systems. Among the various \gls{md}-based methods for thermal conductivity calculations, the \gls{hnemd} method has been proven to be the most efficient one~\cite{2019_prb_fan_hnemd}. In this method, one applies an external driving force 
\begin{equation}
\label{equation:Fe}
\bm{F}_{i}^{\rm ext} = \bm{F}_{\rm e}\cdot\mathbf{W}_{i},
\end{equation}
to create a nonzero heat current. Here $\bm{F}_{\rm e}$ is the driving force parameter with the dimension of inverse length and $\mathbf{W}_{i}$ is the $3\times 3$ per-atom virial tensor (not necessarily symmetric for many-body potentials) \cite{2015_fan_force-gpumd, 2021_prb_Gabourie_Kt, 2021_prb_fan} defined as
\begin{equation} 
\label{equation:virial}
\mathbf{W}_{i} = \sum_{j\neq i} \bm{r}_{ij} \otimes \frac{\partial U_j}{\partial \bm{r}_{ji}},
\end{equation}
where $\otimes$ denotes tensor product between two vectors, $U_j$ is the site energy of atom $j$, and $\bm{r}_{ij}$ is defined as $\bm{r}_j - \bm{r}_i$, $\bm{r}_i$ being the position of atom $i$.
In the linear-response regime, the nonequilibrium ensemble average (denoted by $\langle \rangle$) of the heat current $\bm{J}$ is proportional to the the driving force parameter:
\begin{equation}
\label{equation:J}
\langle J^{\alpha} \rangle =  T V \sum_{\beta} \kappa^{\alpha\beta} F_{\rm e}^{\beta},
\end{equation}
where $T$ is the temperature and $V$ is the volume.
Here the instant heat current is calculated based on the definition \cite{2015_fan_force-gpumd, 2021_prb_Gabourie_Kt, 2021_prb_fan}
\begin{equation} 
\label{equation:J_i}
\bm{J} = \sum_{i} \mathbf{W}_{i} \cdot \bm{v}_{i},
\end{equation}
where $\bm{v}_i$ is the velocity of atom $i$. The thermal conductivity tensor $\kappa^{\alpha\beta}$ can thus be extracted from Eq.~(\ref{equation:J}). The judicious choices of the magnitude of $\bm{F}_{\rm e}$ for the systems here are presented in Fig.~S1 of the \gls{sm}.

Figure~\ref{fig:kappa-time-size-rate}(a) shows the cumulatively averaged $\kappa$ versus time in the \gls{hnemd} simulations for three independent \gls{asi} samples with $N=64,000$ atoms obtained with a quenching rate of $\alpha=10^{11}$ K s$^{-1}$ and a final temperature of $T=300$ K. We see that $\kappa$ converges nicely with the simulation time. For temperatures $T < 100$ K, we use six \gls{asi} samples due to the worse ergodicity in \gls{md} simulations with decreasing temperature. In all the cases, we treat the trajectories for different samples as a whole and divide them into about 100 equally-sized blocks and calculate a proper estimate of the statistical error (measured as the standard error).  Figure \ref{fig:kappa-time-size-rate}(b) presents the distribution of the block-averaged $\kappa$ values.

The \gls{hnemd} method is physically equivalent to the Green-Kubo method and thus has similar finite-size effects as in the Green-Kubo method~\cite{2021_Guxiaokun_jap}, which come from two competing effects~\cite{1986_Ladd_prb}: a finite cell truncates some long-wavelength vibrations and also ignores some scattering events. In disordered materials, the former should dominate and we expect that $\kappa$ will increase with increasing simulation cell size. This is indeed the case for our results shown in Fig.~\ref{fig:kappa-time-size-rate}(c) (see Fig.~S2 in the \gls{sm} for the time convergence of $\kappa$ for each cell size), where we can see that it requires a system size of $N=64,000$ atoms (or a linear size of 11 nm in a cubic cell) to almost converge $\kappa$. This is the size we use for all the subsequent calculations except for those in Sec.~\ref{sec:low_frequency_limit}, where we explore the effects of finite simulation domain size in more depth.

Similar to the structural properties, $\kappa$ in \gls{asi} is also sensitive to the quenching rate $\alpha$, increasing with decreasing $\alpha$ and converging to about $1.82$ W m$^{-1}$ K$^{-1}$ at about $\alpha = 2\times 10^{11}$ K s$^{-1}$, see Fig.~\ref{fig:kappa-time-size-rate}(d) (see Fig.~S3 in the \gls{sm} for the time convergence of $\kappa$ for each quenching rate). That is, a more ordered structure from a lower quenching rate conducts heat better. Based on the  short- and medium-range characterizations, it seems $\alpha = 10^{11}$ K s$^{-1}$ is a safe choice that is also computationally affordable. We thus use this quenching rate in all the subsequent calculations.

It is worth noting that experimentally measured thermal conductivity of \gls{asi} varies from $1.6$ to $4$ W~m$^{-1}~$K$^{-1}$ \cite{Liu2009prl,kown2017ACSnano,1994_Cahill_kappa-expt,2021_prm_Kim_kappa-expt,regner2013nc,2006_prl_zink} at room temperature. This large variation is most likely due to the structural differences of the \gls{asi} samples: more ordered samples tend to have  higher thermal conductivity according to our results above. For example, Liu \textit{et al.} \cite{Liu2009prl} reported a thermal conductivity of 4 W~m$^{-1}$~K$^{-1}$ for a $80$-$\mu$m-thick \gls{asi} film deposited by the hot-wire \gls{cvd} method at room temperature, and they confirmed that their \gls{asi} structures are more ordered and possess higher medium-range order than typical ones. Using another flavor of \gls{cvd}, the vapor-liquid-solid mediated low-pressure \gls{cvd}, Kwon~\textit{et al.} \cite{kown2017ACSnano} also obtained a thermal conductivity up to 4 W~m$^{-1}$~K$^{-1}$ for a $1.7$-$\mu$m-thick \gls{asi} film at $300$~K. On the other hand, \gls{asi} samples prepared by \gls{pvd}, such as sputtering deposition \cite{regner2013nc,1994_Cahill_kappa-expt}, electron-beam deposition \cite{2006_prl_zink} and self-implantation \cite{1999_prb_Khalid,1999_prl_Khalid}, tend to be more disordered. Particularly, the \gls{asi} samples by Zink \textit{et al}. are of high purity without crystallinity \cite{2006_prl_zink}. Our numerical \gls{asi} sampled prepared by the melt-quench-anneal protocol resemble the \gls{pvd} ones \cite{1999_prb_Khalid,1999_prl_Khalid} as has been shown above. Therefore, we will mainly compare our thermal conductivity results against those by Zink \textit{et al} \cite{2006_prl_zink}, which span a large range of temperature (from a few K to a few hundred K).

\subsection{Quantum-statistical correction}

After determining the converged $N$ and $\alpha$, we calculate $\kappa$ for \gls{asi} at different temperatures and compare the results with experimental ones, see Fig.~\ref{fig:kT}. Figure S4 in the \gls{sm} presents the time convergence of $\kappa$ for each temperature. The \gls{hnemd} results (triangles) only agree with the experimental ones around and above room temperature, significantly overshooting at low temperatures.

Considering the fact that \gls{md} simulations follow classical statistics and the relatively high Debye temperature of silicon (about 487~K) \cite{2006_prl_zink}, we expect that this overshooting is mostly due to the missing quantum-statistical effects in the MD simulations. To study this in detail, we note that there is a feasible quantum-correction method based on the spectral thermal conductivity in the \gls{hnemd} formalism~\cite{2019_prb_fan_hnemd}. In this formalism, one can obtain the spectral thermal conductivity $\kappa(\omega,T)$  as a function of the vibrational frequency $\omega$ and temperature $T$ with a Fourier transform of the so-called virial-velocity correlation function~\cite{2019_prb_fan_hnemd, 2021_prb_Gabourie_Kt}, which is a generalization of the spectral heat current approach~\cite{2014_Saaskilahti_prb,2015_Saaskilahti_prb} from interface to bulk materials. The virial-velocity correlation function is defined as \cite{2019_prb_fan_hnemd, 2021_prb_Gabourie_Kt}:
\begin{equation}
    \bm{K}(t) = \sum_i \langle \mathbf{W}_i(0) \cdot \bm{v}_i(t) \rangle.
\end{equation}
The summation is over the atoms in a control volume $V$ of interest, and $\langle  \rangle$ indicates the average over different time origins. The spectral thermal conductivity is then calculated as
\begin{equation}
    \kappa(\omega,T) = \frac{2}{VTF_{\rm e}}\int_{-\infty}^{\infty} \text{d}t e^{\text{i}\omega t} K(t).
\end{equation}
This $\kappa(\omega,T)$ is classical but it can be quantum corrected by multiplying it with a ratio between quantum and classical modal heat capacity~\cite{2016_qc-correction,2016_qc-correction2,2017_nl_qc-correction},
\begin{equation}
\kappa^{\rm q}(\omega,T) = \kappa(\omega,T)\frac{x^2e^x}{(e^x-1)^2},
\label{equation:qc_k_omega}
\end{equation}
where $x=\hbar\omega/k_{\rm B}T$, $\hbar$ is the reduced Planck constant and $k_{\rm B}$ is the Boltzmann constant. The effectiveness of this quantum-correction method for amorphous systems originates from the fact that the population of vibrations has negligible effects on elastic scattering processes.  This is in sharp contrast with crystals, where inelastic phonon-phonon scattering dominates
and the over-populated high-frequency phonons in classical \gls{md} can reduce the lifetime of the low-frequency phonons and a simple correction of the phonon population usually leads to an underestimated thermal conductivity~\cite{Turney2009PRB}. 

\begin{figure}
    \centering
    \includegraphics[width=\columnwidth]{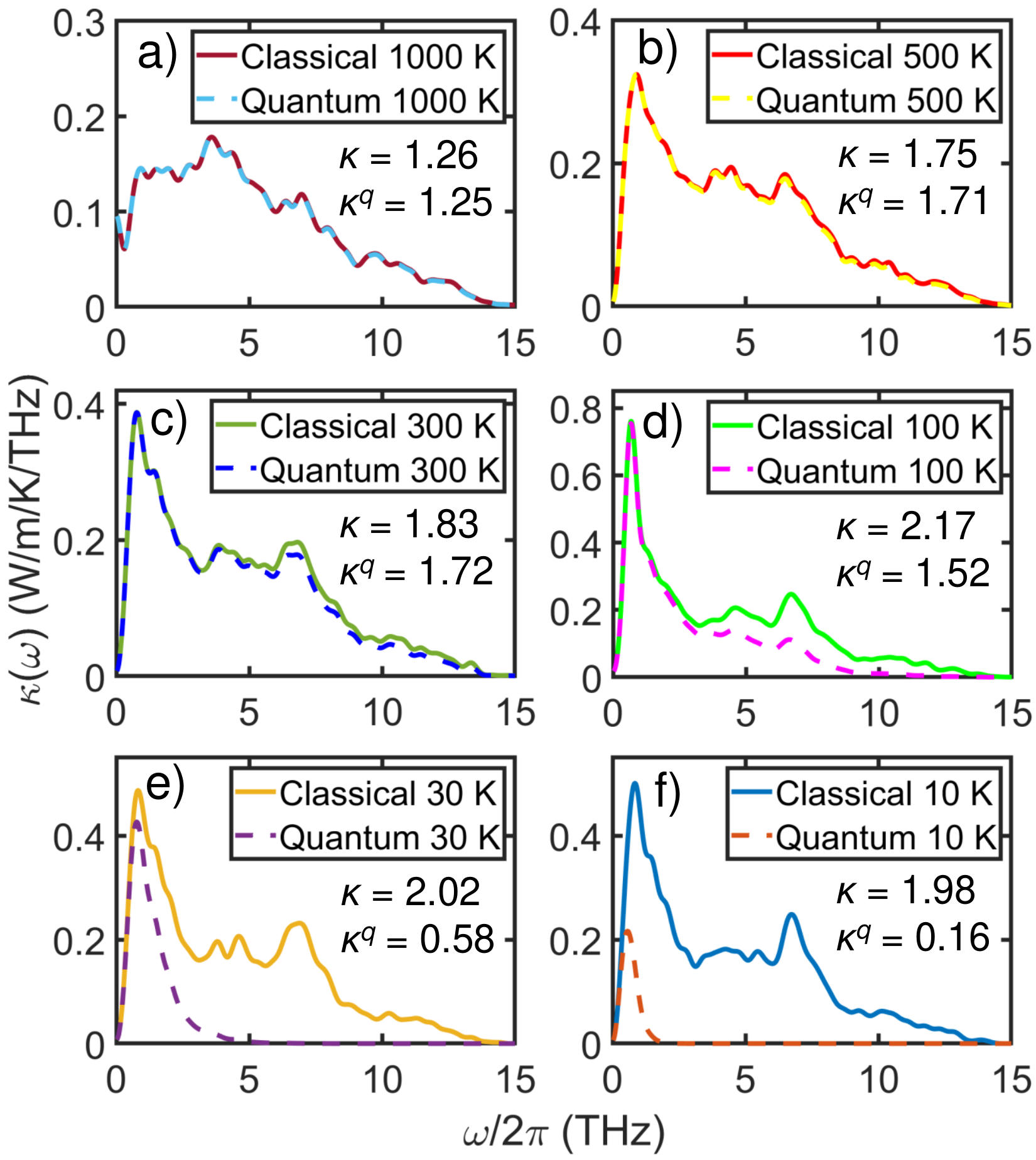}
    \caption{Classical and quantum-corrected spectral thermal conductivity of \gls{asi} at (a) 1000, (b) 500, (c) 300, (d) 100, (e) 30 and (f) 10 K. Integrated thermal conductivity values are indicated in each panel.}
    \label{fig:shc}
\end{figure}

\begin{figure}
\centering
\includegraphics[width=\columnwidth]{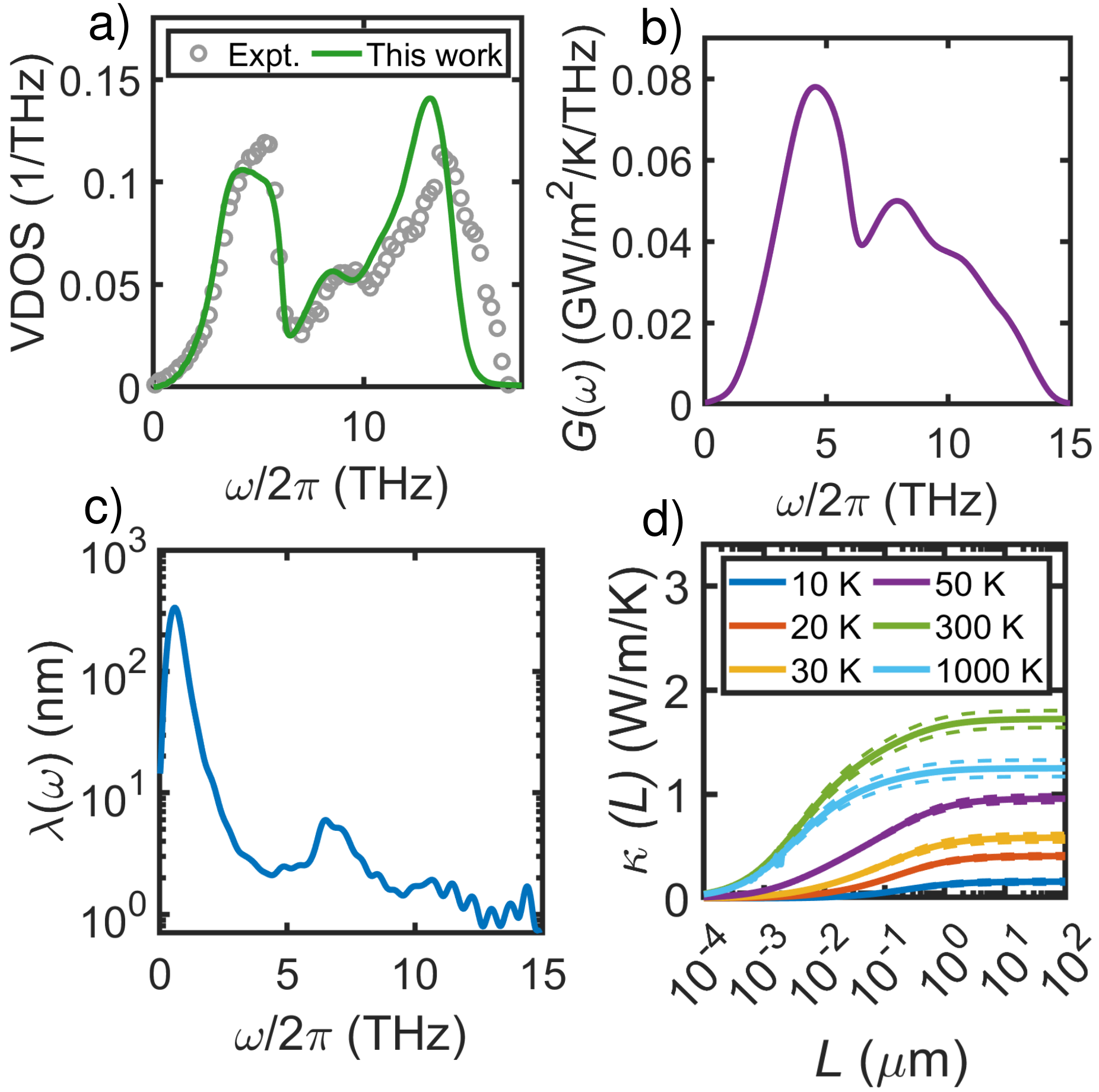}
\caption{(a) Vibrational density of states $\rho(\omega)$, (b) classical ballistic spectral thermal conductance $G(\omega)$ at $20$ K, and (c) vibrational mean free path $\lambda(\omega)$ of \gls{asi} as a function of the vibrational  frequency $\omega/2\pi$ at $20$ K. (d) Quantum-corrected thermal conductivity $\kappa^{\rm q}(L,T)$ as a function of thickness $L$ of \gls{asi} film in the transport direction. Standard errors are given as dashed lines. Experimental values of $\rho(\omega)$ in panel (a) are taken from Kamitakahara \textit{et al.}~\cite{expt_pdos}.}
    \label{fig:gc_lambda_kl_pdos}
\end{figure}

Figure~\ref{fig:shc} shows the classical and quantum-corrected spectral thermal conductivity at different temperatures. Quantum corrections are large at low temperatures and high frequencies, which is consistent with the fact that the populations of the vibrational modes in these conditions are artificially high in classical \gls{md} simulations. The total quantum corrected thermal conductivity $\kappa^{\rm q}(T)$ is then obtained as an integral of $\kappa^{\rm q}(\omega,T)$ over the frequency as 
\begin{equation}
\kappa^{\rm q}(T) = \int_0^{\infty} \frac{\text{d}\omega}{2\pi} \kappa^{\rm q}(\omega,T). 
\end{equation}
At room temperature, $\kappa^{\rm q}(T)/\kappa(T)$ is close to unity at $94\%$, while it becomes as small as $8.1\%$ at 10 K. The strong quantum-statistical effects make the classical \gls{md} results to fail to describe the experimental measurements at low temperatures. After applying the quantum correction, the \gls{hnemd} results (squares in Fig.~\ref{fig:kT}) are much closer to the experimental ones by Zink \textit{et al}~\cite{2006_prl_zink}, but are still slightly too large at the low-temperature limit. We note that if $\kappa(\omega)$ is not available, one may attempt to make a quantum correction based on the \gls{vdos} $\rho(\omega)$:
\begin{equation}
    \tilde{\kappa}^{\rm q}(T) = \kappa(T) \frac{ \int^{\infty}_0 \frac{\text{d}\omega}{2\pi} \rho(\omega)\frac{x^2e^x}{(e^x-1)^2}}
   {\int^{\infty}_0 \frac{\text{d}\omega}{2\pi} \rho(\omega)}.
    \label{equation:qc_rho_omega}
\end{equation}
This is however not quantitatively correct because $\rho(\omega)$ does not contain the information of heat transport that is contained in $\kappa(\omega)$ and weights more for the high-frequency part than $\kappa(\omega)$, as can be seen from a comparison between Fig.~\ref{fig:gc_lambda_kl_pdos}(a) and Fig.~\ref{fig:shc}. Therefore, the quantum correction based on $\rho(\omega)$ results in too small a ratio $\tilde{\kappa}^{\rm q}(T)/\kappa(T)$ as compared to the correct one $\kappa^{\rm q}(T)/\kappa(T)$ from the quantum correction based on $\kappa(\omega)$, as can be clearly seen from Fig.~\ref{fig:kT} (diamond symbols).

\subsection{Length dependence of thermal conductivity}

To understand the overestimation of thermal conductivity in the low-temperature limit using the statistical quantum correction, Eq.~(\ref{equation:qc_k_omega}), we note that the experimental samples are of finite thickness in the transport direction, being $L=130$~nm~\cite{2006_prl_zink}, while $L$ in our \gls{hnemd} simulations should be regarded as infinite. Strong length dependence of $\kappa$ has been experimentally observed in amorphous silicon thin films \cite{braun2016prb} or through \gls{mfp} spectroscopy \cite{pan2020prb}. To enable a more proper comparison with experiments, we need to compute $\kappa$ at a finite $L$. A conventional approach is to perform  heterogeneous \gls{nemd} simulations at different $L$. However, a more computationally efficient and elegant way is to first perform a single \gls{nemd} simulation in the ballistic limit (low $T$ and short $L$) that is equivalent to the atomistic Green's function approach \cite{2019_Li_NEMD}, and then employ the same spectral decomposition method as in \gls{hnemd} \cite{2019_prb_fan_hnemd} to obtain the spectral thermal conductance $G(\omega)$ (see Fig.~\ref{fig:gc_lambda_kl_pdos}(b)):
\begin{equation}
    G(\omega) = \frac{2}{V\Delta T}\int_{-\infty}^{\infty} \text{d}t e^{\text{i}\omega t} K(t).
\end{equation}
Here $\Delta T$ is the temperature difference between the 
heat source and heat sink in the \gls{nemd} setup. 
After this, one can obtain the spectral \gls{mfp} as $\lambda(\omega,T)=\kappa(\omega,T)/G(\omega)$ (see Fig.~\ref{fig:gc_lambda_kl_pdos}(c)), and then obtain the quantum-corrected thermal conductivity at any thickness as (see Fig.~\ref{fig:gc_lambda_kl_pdos}(d))
\begin{equation}
\kappa^{\rm q}(L,T) = \int \frac{\text{d}\omega}{2\pi} \frac{\kappa^{\rm q}(\omega,T)}{1 +\lambda(\omega,T)/L}.
     \label{equation:kl}
 \end{equation}
 
Setting $L=130$ nm as in the experiments~\cite{2006_prl_zink}, our predicted results (circles in Fig.~\ref{fig:kT}) finally agree well with the experimental ones from room temperature down to 10 K. 

In Fig.~\ref{fig:kT}, we also present previous theoretical results by Zhang \textit{et al}.~\cite{zhang2022npj} and Isaeva \textit{et al}.~\cite{isaeva2019nc}. The methods in these works have been well benchmarked against their respective \gls{md} simulations, but the agreement with experimental data is less satisfactory. These two works have used the Stillinger-Weber \cite{stillinger1985prb} and Tersoff \cite{tersoff1989prb} empirical potentials, respectively, indicating that these traditional potentials that reproduce pristine Si may not be accurate enough for a-Si. It would thus be interesting to study if combining these methods and our \gls{nep} model can improve the results, although this is beyond the scope of the current paper.

Finally, to further confirm the reliability of our predictions, we considered possible external stress that can intentionally or accidentally exist in experiments and find that there is no stress-dependence of $\kappa^{\rm q}$ in \gls{asi} (see Fig.~S5 and Fig.~S6 in the \gls{sm} for details).

Note that we have considered temperatures up to 1000 K in our calculations, but as far as we know there are no available experimental data at such high temperatures. Based on our results, $\kappa$ in \gls{asi} at 1000 K is significantly reduced as compared to the room temperature. The validity of this prediction is yet to be confirmed by future experiments. We stress that while the possible contribution to heat conduction by electrons has been ignored in this work it should not be significant below the melting point of \gls{asi}.

\subsection{The low-frequency limit and finite-size effects}
\label{sec:low_frequency_limit}

\begin{figure}
    \centering
    \includegraphics[width=\columnwidth]{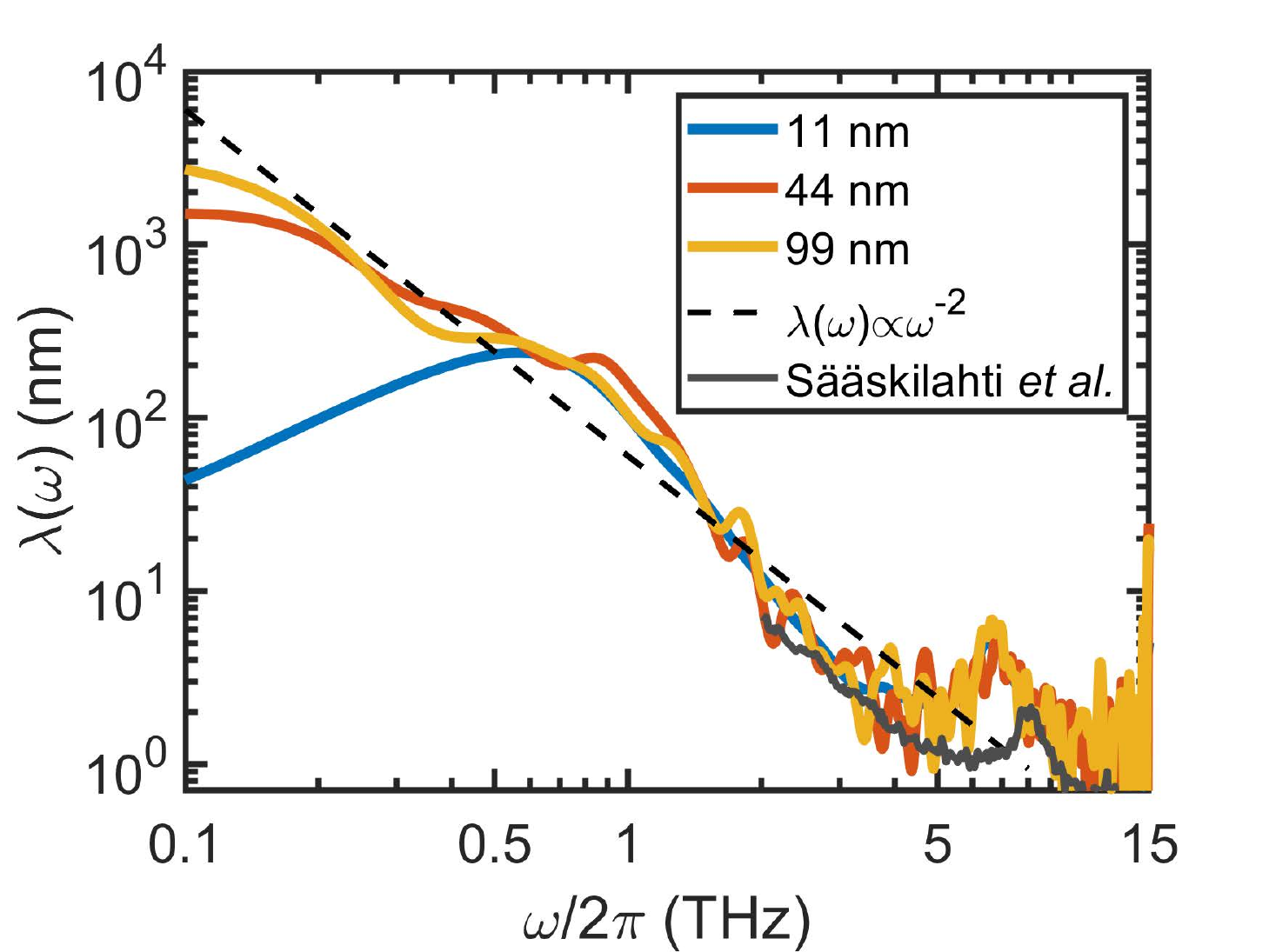}
    \caption{Vibrational mean free path $\lambda(\omega)$ with three simulation domain lengths (from $11$ to $99$ nm) at $300$~K. Previous results from S\"a\"askilahti \textit{et al.} \cite{saaskilahti2016AIPadvances} are added for comparison. See Fig. S7 and Fig. S8 in the \gls{sm} for details on the \gls{hnemd} and spectral decomposition results.}
    \label{fig:mfp_domain_size}
\end{figure}

\begin{figure}
    \centering
    \includegraphics[width=\columnwidth]{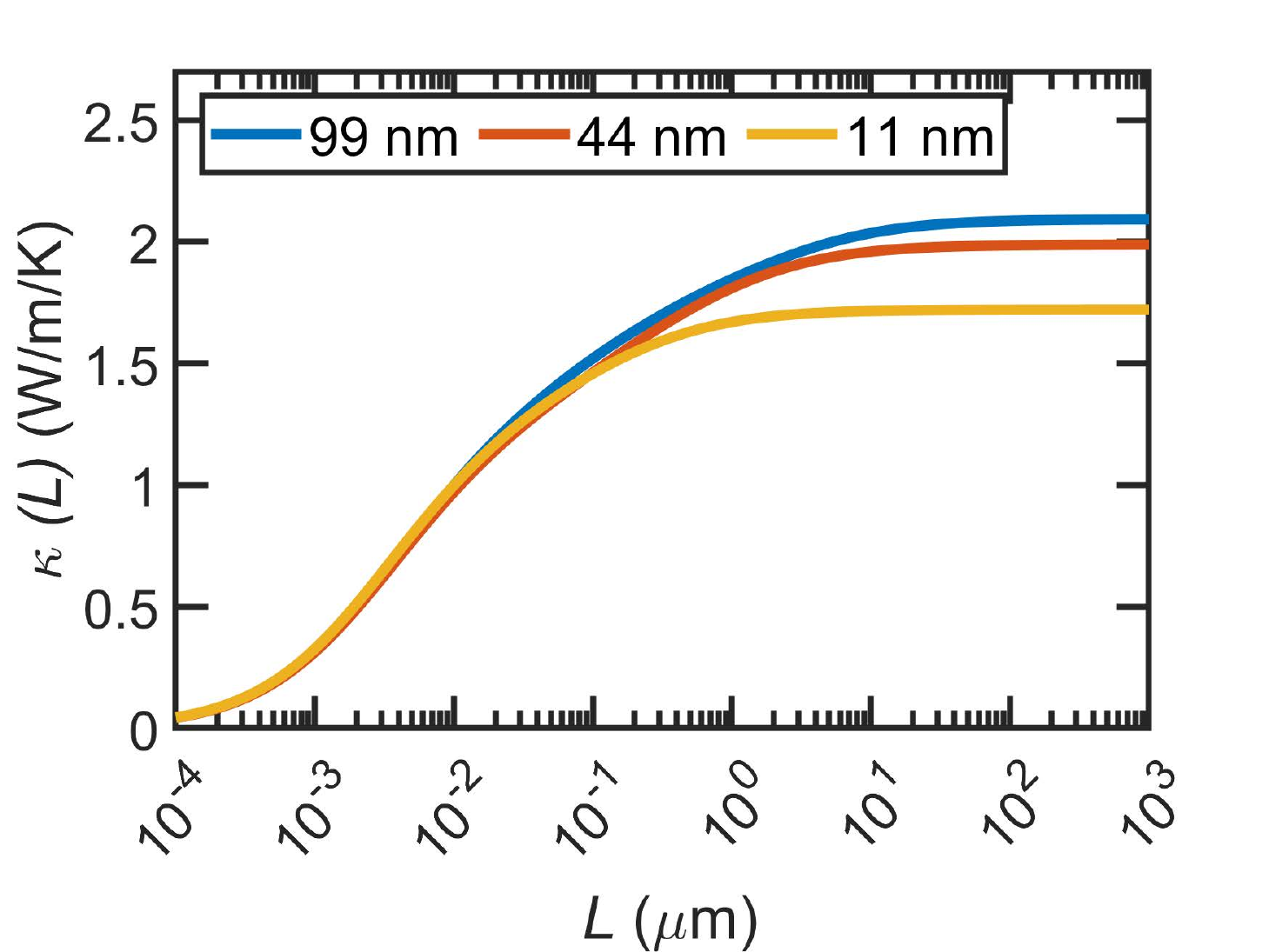}
    \caption{Quantum-corrected thermal conductivity $\kappa^{\rm q}(L)$ of a-Si film at $300$~K as a function of the film thickness $L$ calculated with three simulation domain lengths (11 to 99 nm).}
    \label{fig:kL_domain_size}
\end{figure}

The existence of large vibrational \glspl{mfp} around 1 THz is a manifestation of the existence of propagons \cite{Allen1999pmb}. These propagating phonon-like modes contribute a significant portion to the thermal conductivity, which is consistent with experimental measurements \cite{regner2013nc, braun2016prb, pan2020prb} and previous predictions \cite{Larkin2014prb, Moon2018prb, 2021_zhou_jap, zhang2022npj}. We note that the \glspl{mfp} in Fig.~\ref{fig:gc_lambda_kl_pdos}(c) drops in the low-frequency limit, which is caused by the finite simulation domain length. According to the effective group velocity $v_{\rm g} \approx 8$ km/s of the propagons~\cite{Moon2018prb,Moon2019prm}, a domain length of $D=11$ nm can only support vibrations with frequency down to $v_{\rm g}/D \approx 0.7$ THz. By considering a domain length up to 99 nm, the lower limit of the frequency that can be probed is pushed down to about 0.08 THz, as can be seen in Fig.~\ref{fig:mfp_domain_size}. We also observe that, in the low-frequency limit, the \glspl{mfp} scale as $\lambda(\omega)\propto\omega^{-2}$. However, the \gls{vdos} scales as $\rho(\omega)\propto\omega^2$ (Fig.~\ref{fig:gc_lambda_kl_pdos}(a)), leading to a constant contribution to the thermal conductivity. By increasing the domain length from 11 to 99 nm, the total thermal conductivity is only increased by about 15\%, but the thickness-convergence of the thermal conductivity is extended from about 1 to 10 microns (cf. Fig.~\ref{fig:kL_domain_size}). 

\section{Summary and conclusions}

In summary, we have studied heat transport in \gls{asi} using extensive \gls{md} simulations with an accurate and efficient \gls{mlp} constructed using the \gls{nep} approach. Realistic \gls{asi} samples were first generated using the melt-quench-anneal process. Based on detailed structural analyses and heat transport calculations, we found that both short-range and medium-range structural order increase with reduced quenching rate, and the calculated thermal conductivity accordingly increases. A quenching rate of $10^{11}$~K s$^{-1}$ is determined to be appropriate to generate realistic \gls{asi} samples and converged thermal conductivity. The thermal conductivity calculated from \gls{hnemd} simulations also exhibits notable finite-size effects, requiring a simulation cell with a linear size of 11 nm to reach asymptotic convergence. Based on spectral decomposition techniques, we verified the importance of both quantum statistical effects and a finite sample length commensurate with experiments on the predicted thermal conductivity of \gls{asi}. With a correction to the classical spectral thermal conductivity based on quantum statistics, we obtained good agreement with experiments from 10 K to room temperature using the same sample length. Finally, we also demonstrated that an approximate quantum-correction scheme based on the density of states is inaccurate for a-Si.

\vspace{0.5cm}
\noindent{\textbf{Data availability}}

The inputs and outputs related to the \gls{nep} model training are freely available at the Gitlab repository: \url{https://gitlab.com/brucefan1983/nep-data}. The inputs and outputs of all the \gls{md} simulations are freely available at Zenodo \cite{yanzhou_zenodo}.

\vspace{0.5cm}
\noindent{\textbf{Code availability}}

The source code of \textsc{gpumd} is available at \url{https://github.com/brucefan1983/GPUMD} and the related documentation can be found at \url{https://gpumd.org}.

\begin{acknowledgments}
The authors acknowledge funding from the Academy of Finland,
under projects 321713 (M.A.C. \& Y. W.), 330488 (M.A.C.), 312298/QTF
Center of Excellence program (T.A.-N., Z.F. \& Y.W.), the National Natural Science Foundation of China (NSFC) under grant no. 11974059 (Z.F.), the National Key Research and Development Program of China under grant no. 2021YFB3802100 (P.Q. \& Y.W), and the China Scholarship Council under grant no. CSC202006460064 (Y.W.). The authors also acknowledge computational resources from the Finnish Center for Scientific Computing (CSC) and Aalto University's Science IT project.
\end{acknowledgments}

\bibliography{refs}

\end{document}